\documentclass{ws-ijmpe}

\newcommand{\nuc}[2]{$^{#1}${#2}}

\newcommand{\trans}{{T}}

\begin{document}

\markboth{Bender, Duguet, Heenen, and Lacroix}
         {Regularization of MR EDF calculations}

\catchline{}{}{}{}{}

\title{Regularization of Multi-Reference Energy Density Functional
       Calculations}

\author{M. Bender}
\address{Universit{\'e} Bordeaux, CNRS/IN2P3,
             Centre d'Etudes Nucl{\'e}aires de Bordeaux Gradignan,
             CENBG, Chemin du Solarium, BP120,
             F-33175 Gradignan, France}

\author{T. Duguet}
\address{CEA, Centre de Saclay, IRFU/Service de Physique Nucl{\'e}aire, F-91191 Gif-sur-Yvette, France
\\
National Superconducting Cyclotron Laboratory and Department of Physics and Astronomy,
Michigan State University, East Lansing, MI 48824, USA}

\author{P.-H. Heenen}
\address{PNTPM, CP229,
             Universit{\'e} Libre de Bruxelles,
             B-1050 Bruxelles,
             Belgium}

\author{D. Lacroix}
\address{GANIL, CEA et IN2P3, BP 5027, 14076 Caen Cedex, France}

\maketitle

\begin{history}
\received{(received date)}
\revised{(revised date)}
\end{history}

\begin{abstract}
We report the first application of a recently proposed regularization
procedure for multi-reference energy density functionals, which removes
spurious divergent or non-continuous contributions to the binding energy,
to a general configuration mixing. As an example, we present a calculation
that corresponds to the particle-number and angular momentum projection
of axially symmetric time-reversal invariant quasiparticle vacua of different
quadrupole deformation for the nucleus \nuc{18}{O}. The SIII parameterization
of the Skyrme energy functional is used.
\end{abstract}
%
%
\section{Introduction}

Methods based on energy density functionals (EDF) currently
provide the only set of fully microscopic theoretical tools that can
be applied to all bound atomic nuclei in a systematic manner
irrespective of their mass, isospin, and deformation.\cite{RMP}
Nuclear EDF methods coexist on two distinct levels. On the first level, 
traditionally called "self-consistent mean-field theory" or sometimes 
Hartree-Fock (HF) or Hartree-Fock-Bogoliubov (HFB) method,
a single product state provides the normal and anomalous density
matrices that enter the energy density functional. This
type of method is referred to as a single-reference (SR) approach. 
On the second level,
often called "beyond-mean-field methods", symmetry restoration and
configuration mixing in the spirit of the Generator Coordinate Method
(GCM) can be achieved.\cite{BadHonnef,LesHouches}
At that level, the many-body energy takes the
form of a functional of the transition density matrices that are
constructed from a set of several product states, the number of which
might be on the order of $10^5$ in the most advanced applications. Such a
method, that encompasses the SR one by construction, is referred to
as a multi-reference (MR) approach.

It has been pointed out that MR EDF calculations might be contaminated
by unphysical contributions to the energy.\cite{Taj92a,Ang01a,Dob07a}
In what follows, we provide a summary of the analysis of this
problem and of a regularization scheme to remove it.\cite{I,II,III}
Results for \nuc{18}{O} are used as an illustrative example.
Configuration mixing calculations using projection and GCM techniques
were originally introduced in a Hamilton-operator+wave-function based
framework.\cite{Won75a,BlaRip} It has to be stressed that \emph{none}
of the problems discussed in the present paper appears when such calculations
are performed without making \emph{any} simplifying approximations.

%
%
\section{SR and MR EDF in a nutshell}
In the SR EDF framework, the effective interaction is set-up through an
energy functional
\begin{equation}
\label{eq:ESR}
\mathcal{E}^{\text{SR}}_q
\equiv \mathcal{E}^{\text{SR}}_q [\rho_{qq},\kappa_{qq}, \kappa^{*}_{qq}]
\, ,
\end{equation}
that depends on various local or non-local densities, which themselves are
functionals of the normal and anomalous density matrices of an auxiliary
reference product state $|\text{SR}_q \rangle$ labelled by some collective
coordinate $q$
\begin{equation}
\label{eq:rhoSR}
\mathcal{R}_{qq}
\equiv \left(
  \begin{array}{cc}
  \rho_{qq}       & \kappa_{qq}     \\
  -\kappa_{qq}^*  & 1 - \rho_{qq}^*
  \end{array} \right)
\equiv \left(
  \begin{array}{cc}
  \langle \text{SR}_q | \hat{a}^\dagger \hat{a} |\text{SR}_q \rangle^\trans         & \langle \text{SR}_q |\hat{a} \hat{a} | \text{SR}_q \rangle^\trans     \\
  \langle \text{SR}_q |\hat{a}^\dagger \hat{a}^\dagger | \text{SR}_q \rangle^\trans & \langle \text{SR}_q |\hat{a} \hat{a}^\dagger |\text{SR}_q \rangle^\trans
  \end{array} \right)
= \mathcal{R}_{qq}^2
\, .
\end{equation}
MR EDF calculations rely on the extension of the SR EDF to
non-diagonal energy kernels. There is a set of rules and minimal
requirements\cite{Rob07a} based on symmetry arguments and specific limits
of configuration mixing calculations that is usually agreed on when
constructing the MR EDF. It is common practice to proceed by formal
analogy with the expressions obtained when applying the generalized Wick
theorem\cite{Bal69a} (GWT) to the non-diagonal matrix element of a Hamilton
operator between two product states.\cite{Bon90a}
In such a scheme, the MR EDF corresponding to a
state characterized by a set of quantum numbers $\mu$ becomes a weighted
sum over EDF kernels $\mathcal{E}^{\text{MR}}_{q q'}$ between all possible
combinations of SR states entering the MR calculation
\begin{equation}
\label{eq:EMR}
\mathcal{E}^{\text{MR}}_\mu
= \frac{\sum_{q,q'} f_\mu^* (q)  \,
        \mathcal{E}^{\text{MR}}_{q q'} [\rho_{qq'},\kappa_{qq'}, \kappa^{*}_{qq'}] \,
        f_\mu (q')}
       {\sum_{q'',q'''} f_\mu^* (q'') \,
        \langle \text{SR}_{q''} | \text{SR}_{q'''} \rangle \, f_\mu (q''')}
\, .
\end{equation}
Each EDF kernel $\mathcal{E}^{\text{MR}}_{q q'}$ is constructed
by replacing the density matrices entering the SR EDF
$\mathcal{E}^{\text{SR}}_q$ by their homologue \emph{transition}
density matrices
\begin{equation}
\label{eq:rhoMR}
\mathcal{R}_{qq'}
\equiv \left(
  \begin{array}{cc}
  \rho_{qq'}       & \kappa_{qq'}     \\
  -\kappa_{q'q}^*  & 1 - \rho_{qq'}^*
  \end{array} \right)
\equiv \left(
  \begin{array}{cc}
  \frac{\langle \text{SR}_q | \hat{a}^\dagger \hat{a} | \text{SR}_{q'} \rangle}
       {\langle \text{SR}_q | \text{SR}_{q'} \rangle}^\trans &
  \frac{\langle \text{SR}_q |\hat{a} \hat{a} | \text{SR}_{q'} \rangle}
       {\langle \text{SR}_q | \text{SR}_{q'} \rangle}^\trans     \\
  \frac{\langle \text{SR}_q |\hat{a}^\dagger \hat{a}^\dagger | \text{SR}_{q'} \rangle}
       {\langle \text{SR}_q | \text{SR}_{q'} \rangle}^\trans &
  \frac{\langle \text{SR}_q |\hat{a} \hat{a}^\dagger | \text{SR}_{q'} \rangle}
       {\langle \text{SR}_q | \text{SR}_{q'} \rangle}^\trans
  \end{array} \right)
\end{equation}
and multiplying with the norm kernel
$\langle \text{SR}_q | \text{SR}_{q'} \rangle$
\begin{equation}
\mathcal{E}^{\text{SR}}_q [\rho_{qq},\kappa_{qq}, \kappa^{*}_{qq}]
\to
\mathcal{E}^{\text{MR}}_{q q'} [\rho_{qq'},\kappa_{qq'}, \kappa^{*}_{q'q}]
\, \langle \text{SR}_q | \text{SR}_{q'} \rangle
\, .
\end{equation}
The weights $f_\mu (q)$ entering Eq.~(\ref{eq:EMR}) are either determined
by symmetries that are restored, or by solving the so-called 
Hill-Wheeler-Griffin equation, or by a combination of
both.\cite{LesHouches,Won75a,BlaRip}

\begin{figure}[t!]
\centerline{\psfig{file=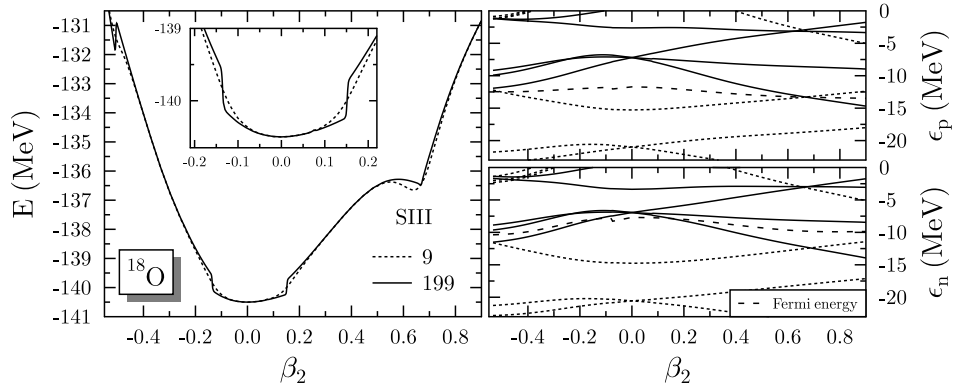,width=12.5cm}}
\caption{\label{eq:pes:spectra}
Left: Particle-number-projected deformation energy surfaces of \nuc{18}{O}
as a function of the axial quadrupole deformation,
one calculated with $L = 9$ discretization points of the integrals
over gauge angles, the other with $L=199$. The inset in the right
panel magnifies the region at small deformation.
Right: corresponding Nilsson diagram of protons (upper right)
and neutrons (lower right). Anomalies in the deformation energy appear when
either a proton or neutron single-particle level crosses the respective
Fermi energy, but they are resolved only when using an excessively
large number of discretization points for the gauge-space integrals.
}
\end{figure}

Over the years it has been realized, however, that in spite of the
many successes of MR EDF calculations the energy functional~(\ref{eq:EMR})
is ill-defined. As an example Fig.~\ref{eq:pes:spectra} shows a
particle-number projected deformation energy curve as a function of
axial mass quadrupole deformation
\mbox{$\beta_2 = \sqrt{5 / 16 \pi} \, (4 \pi/3 R^2 A) \,
\langle \text{SR}_{q} | 2 z^2 - y^2 - x^2 | \text{SR}_{q} \rangle$}
of the underlying SR states, where $R = 1.2 \, A^{1/3} \, \text{fm}$.
At some deformations, the MR energy does
not converge when increasing the number $L$ of discretization points
in the gauge-space integral, although all operator matrix elements
are converged already using 5 points in this case.
Instead, with increasing number of discretization points the energy
curve slowly develops discontinuities, which coincide with the deformations
at which single-particle levels cross the Fermi energy.

First indications for this problem came from an analysis of particle-number
projection that demonstrated that the contribution of
a pair of exactly half-filled levels to the particle-number projected 
energy diverges when direct, exchange and pairing terms do not
recombine in a particular manner.\cite{Ang01a}
The same divergence has been pointed out to appear in approximations
that are tempting to be made for separable forces in a Hamiltonian- and
wave function based framework.\cite{Don98a,Alm01a}
A more recent thorough analysis\cite{Dob07a} in a strict EDF framework
indicates that the divergences are just the
tip of the iceberg of a much larger problem hidden beneath.

One of the key features of all contemporary successful EDFs used
in nuclear structure physics is that in one way or the other the Pauli
principle is sacrificed
for the sake of a simple and efficient description of the in-medium
interaction.\footnote{Some widely used and in general very successful
many-body techniques, such as RPA for example, violate the Pauli-principle
by construction even when a genuine Hamiltonian is used.\cite{BlaRip,RPA}
}
This might concern just a density dependence, or using different
functionals for the particle-hole and particle-particle parts
of the interaction, or neglecting certain (even all) exchange terms,
or any combination of the above. Such a feature of energy functionals
is known to generate spurious ``self-interaction'' in the literature on density
functional theory (DFT) for electronic systems, and there exists a
vast literature on the subject.\cite{SI} All early
analyses\cite{Ang01a,Dob07a,Don98a,Alm01a} point to some violation
of the Pauli principle as a pre\-re\-quisite for the appearance of
the pathologies observed in configuration mixing calculations. The
contamination of the EDF with a spurious self-interaction as such,
however, does \emph{not} lead to the pathologies visible in
Fig.~\ref{eq:pes:spectra}. The problem is that the contribution of
self-interaction (and in addition those of spurious ``self-pairing''\cite{II}
that might appear in calculations with pairing) to the MR EDF
is multiplied with an ill-defined \emph{weight factor} when the MR EDF
is defined in analogy to the GWT along the lines of Eq.~(\ref{eq:EMR}).
This can be shown when constructing the same energy kernels in analogy
to the standard Wick theorem in a suitable basis.\cite{I} Indeed, 
the results inspired from the two Wick theorems
differ in the weight factors that multiply self-interaction and
self-pairing terms. It has to be stressed that there is nothing wrong
with the GWT when used to evaluate matrix elements of operators,
for which there are no such self-interaction terms.

Pathologies introducing discontinuities into MR EDF calculations are easiest
to identify for pure particle-number restoration but they appear for 
\emph{any} configuration mixing. The
gauge-angle integration contained in particle-number projection can be
transformed into a contour integral in the complex plane.\cite{Dob07a,II}
Then, the total energy is proportional to the sum of the residues of
poles at the interior of the integration contour. All operator matrix
elements have only one pole at $z=0$. Two of the pathologies of the MR EDF,
are connected to the appearance of unphysical poles in the EDF
at finite $z^\pm_\mu = \pm i \, |u_\mu|/|v_\mu|$ in the complex
plane, one for each pair of conjugated single-particle
states in the canonical basis.\cite{Dob07a,II} \emph{Divergences} might
appear at certain deformations whenever the integration contour hits
a pole at finite $z^\pm_\mu$. The divergences are accompanied by a
\emph{finite step}, i.e.\ at some deformation the continuation of the
energy surface on one side does not match the energy surface on the
other side. The difference is connected to a pole being either inside
the integration contour (thereby contributing to the energy) or outside
(and thereby not contributing). In rare cases where two or more poles cross
the Fermi energy simultaneously (for example at $\beta_2$ values around
0.7 in Fig.~\ref{eq:pes:spectra}), one observes a sudden change
in the slope of the energy surface instead of a finite step.

Divergences and steps appear whenever the set of MR states contains
at least one pair of orthogonal states,
$\langle \text{SR}_q | \text{SR}_{q'} \rangle = 0$. In particle-number
projection this always happens for half-filled single-particle levels
with $u_\mu^2 = v_\mu^2 = 1/2$ at gauge angle $\pi/2$. There, the
denominator of the transition densities,
Eq.~(\ref{eq:rhoMR}), becomes zero, which explains why the steps in
Fig.~\ref{eq:pes:spectra} coincide with
single-particle levels crossing the Fermi energy. For other
configuration mixings, or when combining particle-number
projection with other mixings, there is no such simple intuitive rule
for the appearance of orthogonal states. However, such situation is known
to appear when mixing quasiparticle vacua with two-quasiparticle
states,\cite{Taj92a} in angular-momentum projection of cranked
HFB states,\cite{Zdu07a} or in combined angular-momentum and
isospin projection,\cite{Sat11a} and the possible appearance of
spurious contributions to the MR EDF has been reported on all
occasions.

The finite steps can appear in MR EDF calculations with any
non-trivial functional, whereas divergences require at least
one term in the EDF that is of higher than second order in normal
and/or anomalous density matrices of a given isospin.\cite{III}
For this reason, there are no divergences towards $\pm \infty$
in Fig.~\ref{eq:pes:spectra}, as the combination of SIII and
a ``volume''-type pairing functional gives a functional that
contains only terms up to second order in each isospin.\cite{II}
Having steps, however, indicates that there are also
spurious contributions to the total energy that are present even 
though no divergence occurs.

There is a third pathology related to the integration contour
hitting branch cuts of the EDF in the complex plane,\cite{Dob07a,III}
which will not be discussed here. It appears when using density
dependencies that become multi-valued functions when extended into the 
complex plane. Again, this might happen for any configuration mixing 
which leads
to complex transition densities. For certain restricted configuration
mixings, this problem can be suppressed using partially projected
densities for the density dependencies instead of transition densities.
This is done, for example, in recent configuration mixing calculations
using the density-dependent Gogny force,\cite{Rod07a} where this strategy
together with the painstaking calculation of \emph{all} exchange
terms suppresses any visible signs of the pathologies discussed here.
Such scheme, however, cannot be expected to work for all imaginable
configuration mixings.\cite{Rob07a}

In projection after variation (PAV) calculations, all of these problems 
are usually hidden as their unambiguous resolution requires a
discretization of the deformation energy surfaces
and of the integrals over gauge (or Euler) angles
that is \emph{much} finer than what is usually used, c.f.\
Fig.~\ref{eq:pes:spectra}. By contrast, a variation after projection
calculation is very much likely to find the divergences.\cite{Dob07a}
But also in a PAV framework, there is no guarantee that just using
low resolution will suppress all consequences of the unphysical
contribution to the MR EDF.

%
%
\section{Regularization of the MR EDF}

A general  method to regularize MR EDF calculations for arbitrary mixing
has been proposed by us.\cite{I} The discussion of the formalism
is beyond the scope of these proceedings, and we refer to the
literature\cite{I,II,IV} for details. Instead, we will give here a
summary of the ideas and key concepts.

As already mentioned, the origin of the divergences and steps
is that postulating the MR EDF as a functional of normal and
anomalous transition densities multiplies contributions to
the EDF that violate the Pauli principle
by ill-defined weight factors. These weight factors turn out to
be different (and well-behaved) when postulating the MR EDF in
analogy with operator matrix elements computed on the basis of the standard
Wick theorem (SWR).\cite{I} As a matter of fact, the weights of 
self-interaction
and self-pairing terms are the \emph{only} ones in the EDF which
are different when comparing a GWT-motivated definition of the
MR EDF with the SWT-motivated one. Obviously the SWT and GWT
are strictly equivalent when evaluating an operator matrix element.

The basic idea of the regularization procedure is to define the
MR EDF in analogy to operator matrix elements computed from the SWT.
However, two technical difficulties arise when trying to do so.
The first one is that the SWT can be applied to the evaluation of
non-diagonal matrix elements only in carefully chosen bases.
By contrast, the GWT can be
used in any basis, or even when using two different bases, one for each
of the two states entering a matrix element,\cite{Bon90a} which of course
explains the GWT's use as the standard procedure in symmetry restoration 
and GCM-type calculations. A basis allowing the use of the SWT
is provided by the canonical basis of the
Bogoliubov transformation between the two quasiparticle bases that
provide the ``left'' and ``right'' product states entering the non-diagonal 
matrix element.\cite{I}
This Bogoliubov transformation between two quasiparticle
bases is not related to pairing correlations, but it has the same
formal properties as the Bogoliubov transformation in HFB theory. The
Bloch-Messiah-Zumino (BMZ) factorization of this Bogoliubov transformation,
which provides the canonical basis that permits to use the SWT, can be
done, but in general turns out to be non-trivial.\cite{IV} 
Pure particle-number projection has
the advantage that this canonical basis can be analytically constructed:
the original ``left'' and ``right'' bases and the canonical one of the
transformation between them differ by state-dependent phase factors only.
This simplification made particle-number projection the testing ground
for first applications of the regularization procedure.\cite{II} Once
a procedure for BMZ factorization of a general Bogoliubov transformation
has been set up, however, it allows for the regularization of any MR EDF
calculation.\cite{IV}

A second difficulty with setting up the MR EDF in analogy to the
SWT is that doing this directly would lead to difficulty in handling
the expression of the functional.
The contributions from specific combinations of single-particle states
have to be taken out from the energy, which prohibits to write the energy
as a functional of one-body densities at all.
A more efficient strategy is to set-up the basic EDF through
one-body transition densities in analogy to the GWT as usual, and to
subtract a \emph{correction} that is defined as the difference
between the expressions obtained when defining the MR EDF in analogy
to the GWT or the SWT, respectively.

\begin{figure}[t!]
\centerline{\psfig{file=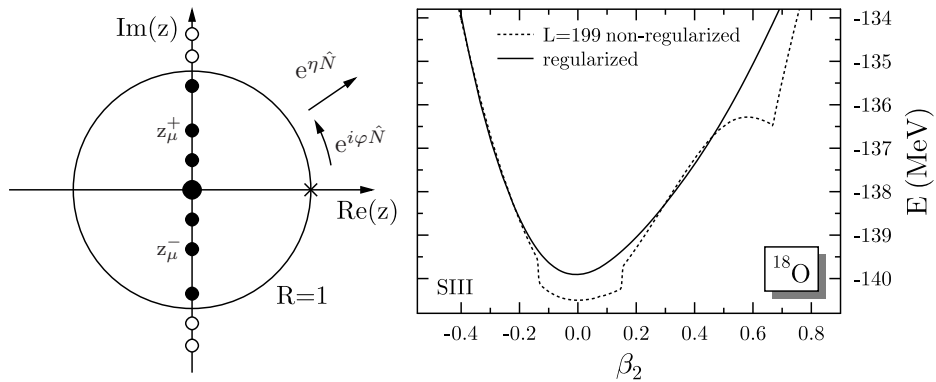,width=12.5cm}}
\caption{\label{eq:pnp}
Left: Schematic view of the analytical structure of the energy kernels
entering the particle-number restored EDF over the complex plane.
Poles marked with filled circles are within the standard circular
integration contour of radius $R = 1$, whereas those outside are
marked with open circles. The cross marks the location of the SR EDF
at $z=1$. The operator $e^{i \varphi \hat{N}}$ produces a rotation in
gauge space, whereas $e^{\eta \hat{N}}$ scales the integration contour.
Right: Non-regularized (dotted line) and regularized (solid line)
particle-number-projected deformation energy for \nuc{18}{O},
calculated with $L = 199$ discretization points of the integrals in
gauge space. The regularized energy curve is independent on $L$ for
$L > 5$, whereas the non-regularized one is not, cf.\
Fig.~\ref{eq:pes:spectra}.
}
\end{figure}

As mentioned above, in the case of pure particle-number restoration,
divergences and steps in the deformation energy surfaces are intimately
connected to the appearance of unphysical poles at
$z^\pm_\mu = \pm i \, |u_\mu|/|v_\mu|$ in the complex plane, see
Fig.~\ref{eq:pnp} for a schematic sketch, and to their evolution with
deformation. The correction, however, does not only (entirely) remove the
contribution of the unphysical poles, but also an unphysical
contribution from the physical pole at $z=0$. The latter is impossible
to identify without having the comparison of SWT- and GWT-based
expressions. Removing only the contribution from the poles at $z^\pm_\mu$
would lead to unphysical results, as their contribution can grow far
beyond any physical scale in the nucleus.\cite{Dob07a,II} It is the
removal of both types of contributions that leads to a meaningful correction;
individually both contributions are very large, of opposite sign, and
nearly cancel. Eventually, the total correction remains smaller than the
energy gain from particle-number restoration as it should.\cite{II}

The regularization, however, is strictly limited to EDFs that
depend on integer powers of the normal and anomalous density matrices
only.\cite{III} This excludes its application to almost all currently
used functionals of acceptable predictive power. One of the few
regularizable Skyrme interactions found in the literature is
SIII,\cite{Bei75a} which we use here for the particle-hole
part of the strong interaction.

As pairing interaction we use a pairing functional of ``volume'' type
of strength 300 MeV fm$^{-3}$.
The widely used Slater approximation to the Coulomb exchange term, however,
falls into the category of multi-valued density-dependent terms.
To obtain a regularizable functional, we keep only the direct term
of the Coulomb energy functional and neglect the approximate exchange
term that was considered in the fit of SIII. As a consequence,
nuclei will be underbound by a few MeV, but this is of no
importance for the purpose of the present article. The
trilinear part of SIII has the particular property that
all its terms are bilinear in densities of one nuclear species
and linear in the other. As an important consequence, there
are no divergences in the non-regularized results shown here, 
only finite steps as already pointed out.

%
%
\section{Results for \nuc{18}{O}}
%
%
\subsection{Pure particle-number restoration}

The comparison of the non-regularized and regularized particle-number
projected deformation energy curve is presented in
Fig.~\ref{eq:pnp} for \nuc{18}{O}. First of all,
the regularization removes  the steps that appeared in the
non-regularized particle-number restored deformation
energy surface of Fig.~\ref{eq:pes:spectra}. For \nuc{18}{O},
it even removes all structure from the deformation energy in the
interval shown, including the pronounced shoulder at $\beta_2 \approx 0.7$.
Second, the correction varies from several hundreds of keVs up to
about 1 MeV depending on the deformation, which is small compared 
to the total binding energy but sometimes accounts for a substantial 
percentage of the energy gain from particle-number restoration.
The regularized EDF converges numerically in the same manner as operator
matrix elements when changing the discretization of MR EDF calculations.
By contrast non-regularized calculations in general do not, and in fact
cannot, converge at all deformations.
The regularized EDF fulfills the sum rules known for particle-number
projected operator matrix elements, whereas the non-regularized EDF
might provide zero and negative particle
numbers with non-zero energies.\cite{II}
The regularized EDF is also \emph{shift invariant} (as are operator matrix
elements), i.e.\ the energy is independent under a $e^{\eta \hat{N}}$
transformation that corresponds to a change of the radius of the
integration contour in the complex plane.\cite{II} By contrast, the
non-regularized EDF might change by \emph{many orders of magnitude}
when varying the radius of the integration contour.\cite{Dob07a,II}
Altogether, this provides strong evidence that the regularized
MR EDF is as well-behaved as operator matrix elements.

%
%
\subsection{Particle-number and angular-momentum restored GCM}

\begin{figure}[t!]
\centerline{\psfig{file=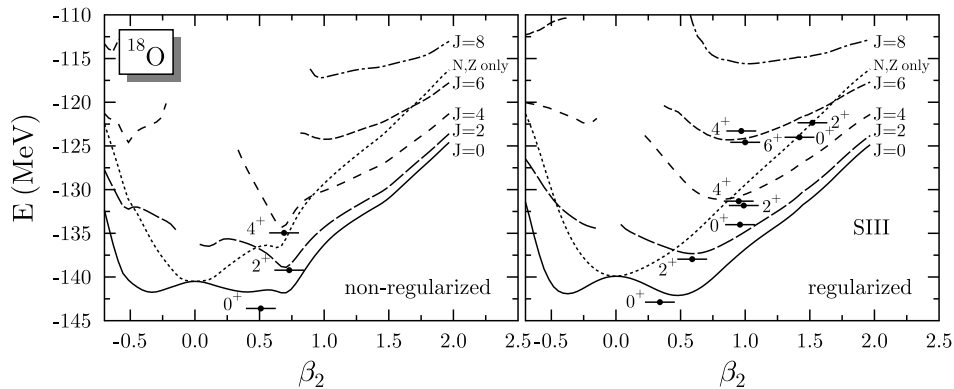,width=12.5cm}}
\caption{\label{eq:gcm}
Left: Non-regularized particle-number and angular momentum projected
deformation energy curves for \nuc{18}{O}, calculated using $L = 9$ gauge
angles and 20 Euler angles, together with the yrast states constructed
by GCM-type mixing of configurations of different deformation
up to $J=4$, plotted at the average deformation of the
intrinsic states they are constructed from.
Right: Regularized particle-number and angular momentum projected
deformation energy curves for \nuc{18}{O}, together with low-lying
states obtained from GCM.
}
\end{figure}

Now we turn to a more complex calculation that combines four different
configuration mixings based on a set of axially symmetric
time-reversal- and space-inversion-invariant quasiparticle vacua:
particle-number restoration of $N=10$ and $Z=8$, i.e.\ the particle
numbers constrained in the underlying SR calculations, angular-momentum
projection, and GCM-type mixing of configurations with different (axial)
shapes.\cite{IV} Results obtained from standard non-regularized
calculations are compared with regularized ones in Fig.~\ref{eq:gcm}.
The shoulder that appears at $\beta_2$ values around 0.7 in the
non-regularized calculations of Fig.~\ref{eq:pnp} leads to a very
localized minimum in the $J=0$, 2 and 4 curves. For higher angular
momenta, the results become irregular. The regularized energy curves,
however, are much smoother.
In the non-regularized calculations, GCM states can be safely constructed
up to $J=4$ only. For higher values of $J$, and non-yrast states in
general, the energies depend too sensitively on the selection of states
entering the GCM, and the Hill-Wheeler-Griffin equation often gives
spurious solutions. These difficulties almost disappear
in the regularized calculations, where the complete low-lying spectrum
can be constructed. It resembles the one of an anharmonic vibrator,
which is in qualitative agreement with the data, where the low-lying
states indeed suggest being such one- and two-phonon multiplets.\cite{O18}
It has to be stressed that in the GCM mixing of axial quasiparticle
vacua for such a small system as \nuc{18}{O} no more than about 8
sufficiently independent intrinsic configurations
(i.e.\ of overlap sufficiently different from 1)
can be found for $J=0$, and even less for higher
values of $J$. This is linked to the very small number of level
crossings in the Nilsson diagrams of Fig.~\ref{eq:pnp}.

%
%

\section{Discussion and Outlook}

We have presented the first application of a regularization scheme
for MR EDF calculation to a general configuration mixing.
The impact of the regularization on the results obtained for
\nuc{18}{O} is quite substantial. It gives ``more regular''
energy curves, and makes the Hill-Wheeler-Griffin equation
much more stable. A much more detailed analysis of regularized
MR EDF calculations of \nuc{18}{O} and other nuclei will be
given in a forthcoming publication.\cite{IV} Already the example
presented here shows that the regularization might become mandatory
in many applications to detailed spectroscopy.

The nucleus \nuc{18}{O} discussed here is a relatively extreme example.
The two single-particle levels crossing simultaneously the Fermi energy
at $\beta_2$ values around 0.7, which dominate many low-lying collective
states, contaminates the most important energy kernels with large
spurious contributions. For all other nuclei we have studied so far,
the overall impact of the regularization on the spectrum of
low-lying states is less dramatic.

Only energy density functionals which are strictly of integer power
in normal and anomalous density matrices are regularizable. The
construction of regularizable functionals of high quality is a
priority for the near future. Eventually, it is likely that additional 
mathematical constraints on the functional form of the MR energy kernel 
must be derived based on group theoretical considerations to make 
symmetry restoration well formulated within the EDF 
context.\cite{Dug10a,Dug11a}

%
%

\end{document}